\documentclass[10pt,conference,final]{IEEEtran}
\IEEEoverridecommandlockouts
\usepackage{amsmath,amssymb,amsfonts}
\usepackage{algorithmic}
\usepackage{graphicx}
\usepackage{textcomp}
\usepackage{xcolor}

\usepackage{tikz}
\usetikzlibrary{quantikz}
\usepackage{makecell}
\usepackage{adjustbox}

\usepackage[backend=bibtex,minbibnames=1, maxbibnames=6, style=ieee, mincitenames=1, maxcitenames=1, url=false]{biblatex}
\usepackage[colorlinks=true,urlcolor=teal,linkcolor=teal,citecolor=teal]{hyperref}

\def\BibTeX{{\rm B\kern-.05em{\sc i\kern-.025em b}\kern-.08em
    T\kern-.1667em\lower.7ex\hbox{E}\kern-.125emX}}

\begin{document}

\title{Variational Quantum Circuits in \\
Offline Contextual Bandit Problems
\thanks{The project this report is based on was supported with funds from the German Federal Ministry for Economic Affairs and Climate Action in the funding program \emph{Quantum Computing – Applications for industry} under project number 01MQ22008B. 
The sole responsibility for the report's contents lies with the authors.}
}

\author{\IEEEauthorblockN{Lukas Schulte}
\IEEEauthorblockA{\textit{Technical University of Munich} \\
Munich, Germany \\
lukas.schulte@tum.de}
\and
\IEEEauthorblockN{Daniel Hein}
    \IEEEauthorblockA{
        \textit{Siemens AG}\\
Munich, Germany \\
hein.daniel@siemens.com}
\and
\IEEEauthorblockN{Steffen Udluft}
    \IEEEauthorblockA{
        \textit{Siemens AG}\\
Munich, Germany \\
steffen.udluft@siemens.com}
\and
\IEEEauthorblockN{Thomas A. Runkler}
    \IEEEauthorblockA{
        \textit{Siemens AG}\\
Munich, Germany \\
thomas.runkler@siemens.com}
}

\maketitle

\begin{abstract}
This paper explores the application of variational quantum circuits (VQCs) for solving offline contextual bandit problems in industrial optimization tasks. Using the Industrial Benchmark (IB) environment, we evaluate the performance of quantum regression models against classical models. Our findings demonstrate that quantum models can effectively fit complex reward functions, identify optimal configurations via particle swarm optimization (PSO), and generalize well in noisy and sparse datasets. These results provide a proof of concept for utilizing VQCs in offline contextual bandit problems and highlight their potential in industrial optimization tasks.
\end{abstract}


\section{Introduction} \label{sec:introduction}

Contextual bandit algorithms have emerged as powerful tools for decision-making under uncertainty. Driven by the increasing demand for personalization and adaptive decision-making, contextual bandits have been widely adopted in various domains, including recommender systems~\cite{liContextualbanditApproachPersonalized2010, gomez-uribeNetflixRecommenderSystem2016}, online advertising~\cite{tangAutomaticAdFormat2013}, and healthcare~\cite{bastaniOnlineDecisionMakingHighDimensional2015}, where decisions must be made based on contextual information to maximize user engagement, click-through rates, or patient outcomes. 

In industrial applications, where systems must be continuously tuned or "steered" for optimal performance, contextual bandits offer a powerful approach to optimizing system configurations.
In these settings, decisions need to be made based on contextual information (\textit{e.g.}, current operational state or environmental conditions), and the overall objective is to maximize some notion of reward (\textit{e.g.}, production throughput, energy efficiency, or product quality).

Applying contextual bandits in industrial environments presents two key challenges: the inherent noise often present in real-world processes and the scarcity of high-quality data due to the cost or time required for sampling. 

Classical machine learning (ML) models have been extensively employed for this purpose~\cite{dudikDoublyRobustPolicy2011, jeunenPessimisticRewardModels2021}. However, the advent of quantum machine learning (QML) has introduced novel opportunities. Specifically, variational quantum circuits (VQCs)---a class of hybrid quantum-classical algorithms that leverage parameterized quantum circuits to approximate complex functions---have demonstrated efficacy in a variety of ML tasks by combining the expressivity of quantum mechanics (\textit{e.g.}, superposition and entanglement) with classical optimization techniques~\cite{mitaraiQuantumCircuitLearning2019}.

This work explores the application of VQCs to solve offline contextual bandit problems, using the Industrial Benchmark (IB) environment as a testbed. The goal is to optimize steering parameters (control inputs) that maximize the expected reward, subject to the constraints of noisy data and unknown process dynamics.  
By leveraging a fixed dataset derived from the IB environment and framing the contextual bandit task as a regression problem, we demonstrate that quantum regression models offer a viable alternative to classical neural networks (NNs) for guiding optimization processes.

In doing so, this work offers a proof of concept for harnessing VQCs in real-world, offline bandit settings and highlights the potential advantages of quantum approaches alongside practical considerations.

Our contributions are as follows:
\begin{enumerate}
\item We propose a hybrid quantum-classical optimization framework combining trained VQC-based reward models with particle swarm optimization (PSO), demonstrating effective identification of superior configurations within the IB environment.
\item We investigate the impact of VQC design choices, including different ansatz designs and data-encoding strategies, and assess their impact on regression accuracy and generalization performance in offline contextual bandit settings.
\item We provide insights into the noise resilience and generalization capabilities of quantum models compared to classical counterparts.
\end{enumerate}

\section{Background and Related Work} \label{sec:background_related_work}

\subsection{Quantum Machine Learning} \label{subsec:quantum_machine_learning}

Parallel to developments in classical ML, the field of QML aims to harness quantum-mechanical effects such as superposition and entanglement to enrich model expressivity and potentially achieve computational advantages~\cite{schuldSupervisedLearningQuantum2018}. Current research in the noisy intermediate-scale quantum (NISQ) era predominantly adopts hybrid quantum-classical approaches, where classical data is processed by VQCs under classical optimization loops. VQCs act as trainable quantum analogs of NNs, featuring parametrized gates arranged in layered circuits~\cite{mitaraiQuantumCircuitLearning2019, chenVariationalQuantumCircuits2020, skolikQuantumAgentsGym2022}. When combined with appropriate data-encoding strategies (\textit{e.g.}, reuploading layers), VQCs can serve as universal function approximators~\cite{lloydQuantumEmbeddingsMachine2020, schuldEffectDataEncoding2021, perez-salinasUniversalApproximationContinuous2024}.

\subsection{Contextual Bandits} \label{subsec:contextual_bandits}

In the field of ML, contextual bandit problems represent a subclass of reinforcement learning (RL) problems that focus on single-step decision-making, where an agent selects actions based on a given context to maximize immediate rewards. This characteristic distinguishes contextual bandits from full RL, which emphasizes long-term planning and cumulative rewards. 

Contextual bandits can be particularly advantageous in industrial optimization settings, where long-term exploration is often infeasible due to high costs, dynamic environments, or safety concerns. Examples include tuning machine settings for turbines or optimizing process control parameters, where identifying optimal configurations under limited data availability is crucial.

In such scenarios, contextual bandits are often utilized in an \emph{offline} setting, where the problem reduces to supervised learning of reward predictions from historical context-action pairs~\cite{fosterPracticalContextualBandits2018, metevierOfflineContextualBandits2019, simchi-leviBypassingMonsterFaster2021, brandfonbrenerOfflineContextualBandits2021}. Although this approach mitigates some practical barriers, challenges such as distributional shift and overfitting remain~\cite{nguyen-tangOfflineNeuralContextual2022}. Despite these obstacles, recent advances suggest that large or flexible models can still perform well under mild assumptions, a phenomenon sometimes referred to as "benign overfitting"~\cite{simchi-leviBypassingMonsterFaster2021}. This observation naturally leads to the question of whether quantum models, with their unique properties and potential advantages, could offer benefits in this offline contextual bandit setting.

\subsection{Quantum Approaches to Contextual Bandits} \label{subsec:vqcs_contextual_bandits}

Offline contextual bandits are known to be reducible to supervised learning. Therefore, quantum models have the potential to replace classical regressors in the context of reward prediction. Prior work has shown that VQCs can match or exceed classical NNs in certain function-approximation tasks, sometimes using fewer parameters or leveraging quantum-specific inductive biases~\cite{duExpressivePowerParameterized2018, chenVariationalQuantumCircuits2020}. However, their application to offline contextual bandit problems has been comparatively underexplored.

Existing quantum approaches to bandit problems have primarily focused on fully quantum scenarios or quantum algorithms addressing online settings~\cite{huTraining2019,Lumbreras2022multiarmedquantum}. \citeauthor{casaleQuantumBandits2020}~\cite{casaleQuantumBandits2020} and \citeauthor{Wang_You_Li_Childs_2021}~\cite{Wang_You_Li_Childs_2021} developed quantum algorithms for the online best-arm identification problem, leveraging quantum amplitude amplification to achieve quadratic speedups. Similarly, \citeauthor{brahmachariQuantumContextualBandits2024}~\cite{brahmachariQuantumContextualBandits2024} proposed an online contextual bandit framework specifically tailored to quantum data, focusing on recommending quantum states given quantum observables as contexts. 

In contrast, our work explicitly addresses the offline contextual bandit setting, where models learn exclusively from pre-collected classical datasets without interactive exploration. 
By employing VQCs as regression models on classical data, our approach focuses on generalization, robustness to data scarcity and noise, and optimization capability, distinguishing itself significantly from prior quantum bandit research. 

This gap in exploring VQCs for offline contextual bandit problems motivates our investigation into whether quantum models, trained on offline industrial data, can guide decision-making as effectively or better than classical baselines.

\section{Methodology} \label{sec:methodology}

\subsection{The Industrial Benchmark Environment} \label{subsec:ib_environment}

The IB environment models complex industrial processes with high-dimensional state spaces, nonlinear dynamics, and stochastic behavior typically found in industrial processes. 
It aims to capture real-world industrial applications, such as finding the most optimal valve settings for gas turbines or the most optimal pitch angles and rotor speeds for wind turbines. While typically used for RL, we adapt it here to an offline contextual bandit setting.

At each time step $t$, the agent takes an action $a_t \in [-1,1]^3$ representing changes to three steering variables: velocity $v$, gain $g$, and shift $h$. These steering variables are bounded to $[0,100]$.
For detailed equations governing the update mechanism of these steering variables, we refer the reader to the original description provided in~\cite{heinBenchmarkEnvironmentMotivated2017}.

The environment is characterized by a setpoint $p$, representing an external influencing factor that acts as the context in our contextual bandit formulation. This setpoint remains fixed regardless of the agent's actions but influences the environment's dynamics. It can be thought of as a required load in a power plant, which affects other dynamics such as the fatigue $f_t$ and the consumption $c_t$ of resources within the system. 
These combine to determine the reward $r_{t+1}$ for transitioning from state $s_t$ to $s_{t+1}$ by performing action $a_t$ according to
\begin{equation}
 r_{t+1} = -c_{t+1} - 3f_{t+1}\,,
\end{equation}
with consumption $c_{t+1}$ and fatigue $f_{t+1}$ weighted in a 1:3 ratio.

For our contextual bandit framework, we collect a static dataset of interactions using the OpenAI Gym wrapper for the IB environment\footnote{\url{https://github.com/siemens/industrialbenchmark/tree/master}}.
We employ an offline data-collection procedure by discretizing $p$, $v$, $g$, $h$ to $\{0,10,\dots,100\}$, yielding $11^4=14{,}641$ unique combinations of setpoints ($p$) and steering variables ($v, g, h$). For each grid point:

\begin{enumerate}
    \item Initialize a new environment with setpoint $p$ and a random seed, then apply actions to move ($v_t,g_t,h_t$) toward the desired operating point ($v,g,h$). These intermediate actions and states within the IB are not part of our final dataset.
    \item Apply a zero-action $[0,0,0]$ for $100$ time steps, once ($v,g,h$) is reached, as a swing-in phase to allow transient dynamics to stabilize.
    \item Average the reward over $100$ time steps to obtain the final fitness value for the grid point. This averaged reward becomes the target variable, $y$, in our dataset.
\end{enumerate}

This data collection and transformation approach condenses dynamic transitions into static snapshots of operating points, effectively collapsing the system's temporal dynamics into an input-output mapping $\mathbf{x}=(p,v,g,h)\mapsto y$. 
Consequently, the contextual bandit problem reduces to a supervised regression task, where $p$ serves as the context and $v, g, h$ represents the actions, with $y$ representing the immediate reward. The objective then is to learn a reward model (RM) that accurately predicts the reward for each action given the context~\cite{fosterPracticalContextualBandits2018, simchi-leviBypassingMonsterFaster2021}. The inherent noise and sparsity of this dataset, due to the limited grid sampling and the stochasticity of the IB, create significant challenges for training accurate regression models.

\begin{figure}[tb]
    \centerline{\includegraphics[width=0.75\columnwidth]{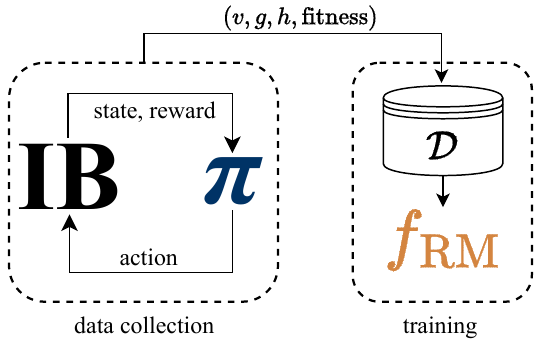}}
    \caption{The IB environment is discretized into a grid with $11^4=14{,}641$ data points, representing combinations of setpoints $(p)$ and steering variables $(v, g, h)$. Rewards at each point are averaged after stabilization, transforming dynamic transitions into static snapshots for training a regression-based reward model.}
    \label{fig:data_collection}
\end{figure}

Fig.~\ref{fig:data_collection} visualizes this data collection process. For additional details, Fig.~\ref{fig:ib_fitness_dist} in Appendix~\ref{app:ib_env} shows the distribution of fitness values across different setpoints, with corresponding statistical summaries provided in Table~\ref{tab:fitness_statistics}.

\subsection{Quantum Regression Model Setup} \label{subsec:quantum_regression}

We consider a regression task mapping a four-dimensional input $\bigl(p, v, g, h\bigr)$ (representing a setpoint and steering variables in the environment) to a scalar fitness (or reward). Real-world data is typically noisy and sparsely sampled, complicating both noise handling and interpolation beyond observed grid points.

The quantum models using VQCs to approximate the reward function are constructed by integrating three key elements, as described by Schuld and Petruccione~\cite{schuldMachineLearningQuantum2021}:
\begin{enumerate}
    \item A data encoding circuit $S(x)$ embedding classical inputs into a quantum state via Pauli-X rotations.
    \item A parameterized quantum circuit defined by a specific ansatz $W(\theta)$.
    \item A measurement operator $\mathcal{M}$ yielding an expectation value of the prepared quantum state after applying the unitary $U(\theta) = W(\theta)S(x)$ to the initial state $\ket{0}$.
\end{enumerate}

In its most general form, the circuit can, therefore, be described as being composed of an embedding layer $S(x)$ and a parametrized block $W(\theta)$
\begin{equation}
 U(x, \theta) = W(\theta) S(x)\,,
\end{equation}
where $x$ is the input data and $\theta$ are the trainable parameters of the circuit.
The model is then defined by the function
\begin{equation}
 f(x, \theta) = \bra{\phi(x, \theta)} \mathcal{M} \ket{\phi(x, \theta)}\,,
\end{equation}
where $\ket{\phi(x, \theta)} = U(x, \theta)\ket{0}$ is the quantum state prepared by the quantum circuit when applied to the initial state $\ket{0}=\ket{0 \ldots 0}$.

\subsubsection{Data Encoding} \label{subsubsec:data_encoding}

We employ angle encoding with Pauli-X rotations to embed classical inputs into a quantum state, as it balances model expressiveness and circuit depth~\cite{chenVariationalQuantumCircuits2020, skolikQuantumAgentsGym2022}. In essence, each feature $x_i$ is encoded by applying $R_x(x_i)$ to the initial qubit state $\ket{0}$. This approach has been shown to broaden the accessible frequency spectrum of the VQC, enabling the model to represent more complex functions~\cite{schuldEffectDataEncoding2021, ostaszewskiStructureOptimizationParameterized2021}.

By repeating the encoding operation---either via data reuploading~\cite{perez-salinasDataReuploadingUniversal2020} or parallel encoding~\cite{rebentrostQuantumSupportVector2014, mitaraiQuantumCircuitLearning2019}---we further expand the frequency spectrum, thus improving the circuit's approximation power. In addition, integrating trainable scaling weights within these rotations allows the circuit to adapt the effective input range dynamically, addressing the risk of "frequency mismatch" and enhancing model performance on real-world data~\cite{lloydQuantumEmbeddingsMachine2020, jerbiParametrizedQuantumPolicies2021}.

\subsubsection{Variational Circuit Ansatz} \label{subsubsec:ansatz}

In this study, we investigate two VQC ansatz designs. 
The first design is referred to as a hardware-efficient ansatz (HEA), which was first introduced by \citeauthor{kandalaHardwareefficientVariationalQuantum2017}~\cite{kandalaHardwareefficientVariationalQuantum2017}. This ansatz involves alternating single-qubit rotations and controlled-Z gates, which are known to be hardware-efficient. 
Secondly, an alternating combination of Circuit 11 and Circuit 9 is employed, adapted from \citeauthor{simExpressibilityEntanglingCapability2019a}~\cite{simExpressibilityEntanglingCapability2019a}, where the expressibility and entangling capability of different ansatz designs were compared. 
The combination of these two ansatz designs is motivated by \citeauthor{steinBenchmarkingQuantumSurrogate2024a}~\cite{steinBenchmarkingQuantumSurrogate2024a}, which demonstrated that the joint application of these two designs yields good results in terms of expressibility and entangling capability while maintaining an effective ratio of circuit parameters to encoding operations.

\begin{figure*}[tb]
    \centerline{
    \begin{adjustbox}{width=0.8\textwidth}
        \begin{quantikz}[font=\scriptsize, column sep=18pt, row sep={18pt,between origins}]
            \lstick{\ket{0}} & \gate{R_x(x_0)}\gategroup[wires=4,steps=1,style={dashed, rounded corners, inner xsep=0pt}, background]{{\scshape fm}} & \gate{R_x(\theta_0)}\gategroup[wires=4,steps=3,style={dashed, rounded corners, inner xsep=0pt}, background]{{\scshape variational layer}} & \gate{R_y(\theta_1)} & \gate{R_z(\theta_2)} & \ctrl{1}\gategroup[wires=4,steps=4,style={dashed, rounded corners, inner xsep=0pt}, background]{{\scshape entangling layer}} & \qw & \qw & \ctrl{} & \qw \: \; \ldots \: \; & \meter{}\gategroup[wires=4,steps=1,style={dashed, rounded corners, inner xsep=0pt}, background]{{\scshape msmt}} \\
            \lstick{\ket{0}} & \gate{R_x(x_1)} & \gate{R_x(\theta_3)} & \gate{R_y(\theta_4)} & \gate{R_z(\theta_5)} & \ctrl{} & \ctrl{1} & \qw & \qw & \qw \: \; \ldots \: \; & \meter{} \\
            \lstick{\ket{0}} & \gate{R_x(x_2)} & \gate{R_x(\theta_6)} & \gate{R_y(\theta_7)} & \gate{R_z(\theta_8)} & \qw & \ctrl{} & \ctrl{1} & \qw & \qw \: \; \ldots \: \; & \meter{} \\
            \lstick{\ket{0}} & \gate{R_x(x_3)} & \gate{R_x(\theta_9)} & \gate{R_y(\theta_{10})} & \gate{R_z(\theta_{11})} & \qw & \qw & \ctrl{} & \ctrl{-3} & \qw \: \; \ldots \: \; & \meter{}
        \end{quantikz}
    \end{adjustbox}}
    \caption{Visualization of a single layer of the first circuit architecture used, consisting of a feature map layer, a variational layer, and an entanglement layer, based on the HEA.}
    \label{fig:circuit_1}
\end{figure*}
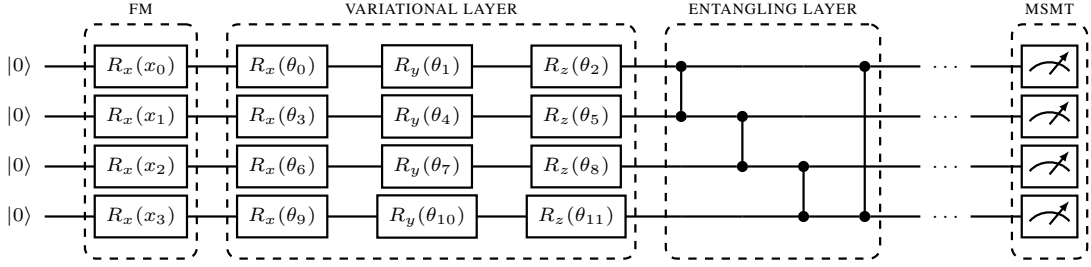

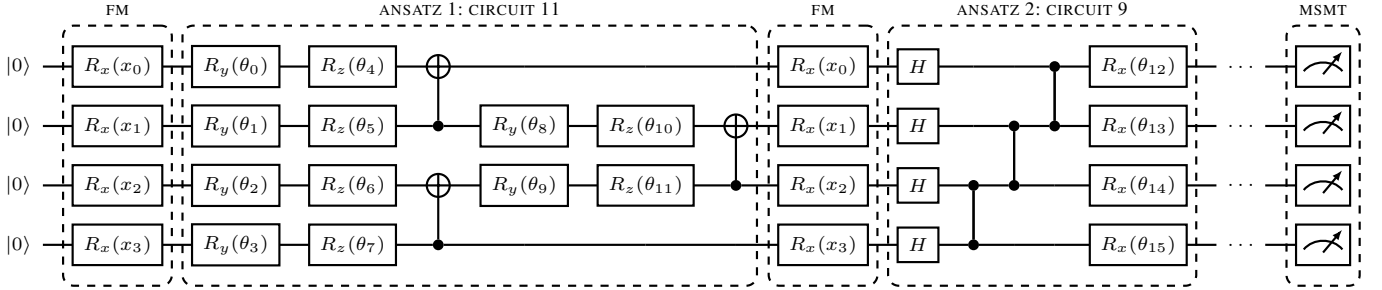
\begin{figure*}[tb]
    \centerline{
    \begin{adjustbox}{width=1.0\textwidth}
        \begin{quantikz}[font=\scriptsize, column sep=11pt, row sep={22pt,between origins}]
            \lstick{\ket{0}} & \gate{R_x(x_0)}\gategroup[wires=4,steps=1,style={dashed, rounded corners, inner xsep=0pt}, background]{{\scshape fm}} & \gate{R_y(\theta_0)}\gategroup[wires=4,steps=6,style={dashed, rounded corners, inner xsep=0pt}, background]{{\scshape ansatz 1: circuit 11}} & \gate{R_z(\theta_4)} & \targ{} & \qw & \qw & \qw & \gate{R_x(x_0)}\gategroup[wires=4,steps=1,style={dashed, rounded corners, inner xsep=0pt}, background]{{\scshape fm}} & \gate{H}\gategroup[wires=4,steps=5,style={dashed, rounded corners, inner xsep=0pt}, background]{{\scshape ansatz 2: circuit 9}} & \qw & \qw & \ctrl{1} & \gate{R_x(\theta_{12})} & \qw \: \; \ldots \: \; & \meter{}\gategroup[wires=4,steps=1,style={dashed, rounded corners, inner xsep=0pt}, background]{{\scshape msmt}}  \\
            \lstick{\ket{0}} & \gate{R_x(x_1)} & \gate{R_y(\theta_1)} & \gate{R_z(\theta_5)} & \ctrl{-1} & \gate{R_y(\theta_8)} & \gate{R_z(\theta_{10})} & \targ{} &  \gate{R_x(x_1)} & \gate{H} & \qw & \ctrl{1} & \ctrl{-1} & \gate{R_x(\theta_{13})} & \qw \: \; \ldots \: \; & \meter{}  \\
            \lstick{\ket{0}} & \gate{R_x(x_2)} & \gate{R_y(\theta_2)} & \gate{R_z(\theta_6)} & \targ{1} & \gate{R_y(\theta_9)} & \gate{R_z(\theta_{11})} & \ctrl{-1} &  \gate{R_x(x_2)} & \gate{H} & \ctrl{1} & \ctrl{-1} &\qw & \gate{R_x(\theta_{14})} & \qw \: \; \ldots \: \; & \meter{}  \\
            \lstick{\ket{0}} & \gate{R_x(x_3)} & \gate{R_y(\theta_3)} & \gate{R_z(\theta_7)} & \ctrl{-1} & \qw & \qw & \qw &  \gate{R_x(x_3)} & \gate{H} & \ctrl{-1} & \qw & \qw & \gate{R_x(\theta_{15})} & \qw \: \; \ldots \: \; & \meter{}  
        \end{quantikz}
    \end{adjustbox}}
    \caption{Visualization of two layers of the second circuit architecture used, consisting of feature map (FM) layers and alternating variational layers based on the Circuit 11 and Circuit 9 designs from~\cite{simExpressibilityEntanglingCapability2019a}, followed by a standard measurement operation across all qubits (denoted by "MSMT").}
    \label{fig:circuit_2}
\end{figure*}

Both approaches include the feature-map layer ($R_x$ encodings), variational parameters (\textit{e.g.}, rotations about x,y,z), and a measurement in the Pauli-Z basis on all qubits. Additional scaling weights at the measurement operator can further tune the output range~\cite{jerbiParametrizedQuantumPolicies2021}.
The VQC models are implemented using TensorFlow Quantum~\cite{broughtonTensorFlowQuantumSoftware2021}, a Python library that integrates quantum circuits with classical ML frameworks. 
The circuits are visualized in Fig.~\ref{fig:circuit_1} and Fig.~\ref{fig:circuit_2}.

\subsubsection{Measurement and Loss} \label{subsubsec:measurement_loss}

Quantum models are simulated via TensorFlow Quantum \cite{broughtonTensorFlowQuantumSoftware2021} on classical hardware, measuring the expectation value of Pauli-Z operators for regression outputs. The loss function is defined as the mean squared error (MSE) between predicted and true rewards for each action-context pair.

\subsubsection{Classical Baselines} \label{subsubsec:classical_baselines}

For comparison, we train feed-forward NNs with a range of depths and parameter counts, ensuring approximate parity with the VQCs. Alongside standard activations (ReLU), we also test a sine activation, leveraging the periodic nature of the rotational angles \cite{sitzmann2020implicitneuralrepresentationsperiodic}. 
Using these classical baselines, we can asses when quantum coding offers an advantage over established NN architectures.
The classical baselines are constructed using TensorFlow~\cite{tensorflow2015-whitepaper} and Keras~\cite{chollet2015keras}. 

\subsection{Data Preprocessing} \label{subsec:data_preprocessing}

Our dataset $\mathcal{D} = \{(x_i, y_i)\}_{i=1}^{N}$ includes $N=14{,}641$ samples, each $x_i=(p_i, v_i, g_i, h_i)$ drawn from $\{0, 10, \ldots, 100\}^4$ and a negative scalar reward $y_i$. We apply three main preprocessing steps:

\begin{enumerate}
    \item \textbf{Input Scaling:} Each feature (setpoint or steering variable) is normalized to intervals such as [-1,1] or [0,1] to align with the periodic domain of rotational quantum gates~\cite{schuldEffectDataEncoding2021}. 
 Formally, $x^{\prime}_{i,j} = \frac{x_{i,j} - \min(x_j)}{\max(x_j) - \min(x_j)} \cdot (b - a) + a$, where $\{a,b\}$ defines the target interval.
    \item \textbf{Target Scaling:} Because the raw (negative) rewards exhibit outliers and skew, we apply a sign-preserving log transform and subsequently rescale them to $[-0.5,0.5]$. This compressed range fits well with the $\sigma_Z$ measurement operator, whose outputs lie in $[-1,1]$, thus improving training stability~\cite{bergholmPennyLaneAutomaticDifferentiation2022}.
    \item \textbf{Encoding Scaling Weights:} We allow trainable scaling factors $w_{d_j}$ for each feature, so that $x^{\prime\prime}_{i,j} = w_{d_j} \,x^{\prime}_{i,j}$, before applying the Pauli-X rotation $R_x(x^{\prime\prime}_{i,j})$. Empirical studies show that this adaptive rescaling can enhance expressiveness in shallow circuits~\cite{jerbiParametrizedQuantumPolicies2021}.
\end{enumerate}

\subsection{Optimization Procedure} \label{subsec:optimization_procedure}

\subsubsection{Particle Swarm Optimization} \label{subsubsec:pso}

To address the contextual bandit problem, we use PSO to optimize the steering variables ($v, g, h$) for each context ($p$). PSO explores the continuous action space using the trained VQC-based reward model as a surrogate, iteratively refining configurations to find superior configurations, relative to the best contained in the dataset, that lie between the discrete grid points that are present in the original dataset. The optimization employs Nevergrad~\cite{bennetNevergradBlackboxOptimization2021} with a swarm size of $40$ and a budget of $1000$ function evaluations per context. 

\subsubsection{Ground Truth Evaluation and Relative Optimization Gain} \label{subsubsec:ground_truth_evaluation_ROG}

To validate the configurations identified in the PSO optimization process, they are evaluated in the IB gym environment to measure their true performance. This involves initializing the IB simulator with the setpoint and transitioning the system to the target configuration found by PSO. 
To ensure robust performance estimates, 1000 trajectories are generated for each candidate configuration, with the reward signal averaged across these trajectories to provide an empirical ground truth performance. 
The same evaluation process is applied to the best configuration in the original dataset for each setpoint, ensuring a fair comparison between PSO-identified solutions and the dataset baseline.

To quantify the effectiveness of the optimization process, the Relative Optimization Gain (ROG) metric is introduced. 
ROG measures the cumulative improvement in ground truth values between the configurations identified by PSO and the best configurations in the dataset across the setpoints $\mathcal{P} = \{0, 10, 20, \dots, 100\}$ present in the dataset. 
For a given setpoint $p \in \mathcal{P}$, let $GT_{\text{PSO}}(p)$ and $GT_{\text{DB}}(p)$ denote the ground truth values of the configurations found by the PSO process and the best configuration in the dataset, respectively.

The aggregated metric over all setpoints is computed as the cumulative difference
\begin{equation}
 \text{ROG} = \sum_{p \in \mathcal{P}} \left(GT_{\text{PSO}}(p) - GT_{\text{DB}}(p)\right)\,.
\end{equation}

A positive ROG value indicates that the PSO-identified configurations demonstrate superior performance in terms of ground truth values in comparison to the established dataset baseline. This comprehensive evaluation and metric provide a robust foundation for assessing the practical performance gains achieved by the optimization process, thereby demonstrating its efficacy in bridging the discrepancy between offline model predictions and real-world system behavior.

\section{Experimental Setup} \label{sec:experimental_setup}

\subsection{Implementation Details} \label{subsec:implementation_details}

All models use the same 70/15/15 split for the training, validation, and test sets, respectively. 
The training process is conducted using a gradient-based optimization algorithm with a learning rate of $0.01$ while using a learning rate scheduler to reduce the learning rate by a factor of $0.5$ if the validation loss does not improve for $7$ epochs. The training is conducted for a maximum of $100$ epochs, with early stopping applied if the validation loss does not improve for $10$ epochs. The model weights, which were updated using the training set according to the loss function and performed best on the validation set, are restored as the final model weights. The models are then evaluated on the test set to assess their generalization performance. 

\subsection{Hyperparameter Grids} \label{subsec:hyperparameter_grids}

We systematically search over a grid of 1152 hyperparameter configurations for both quantum and classical models, spanning architectural design, training configuration, and data handling strategies. Each dimension in this hyperparameter "hypercube" (\textit{e.g.}, layer counts, activation functions, batch sizes, etc.) is systematically varied to explore 1152 configurations per model type. For classical models, we adjust network depth (1-4 layers), layer widths (40-128 neurons), activation (ReLU, tanh, sin), optimizers (Adam, SGD), input/output scalings, data stratification, and sample weighting. 

In quantum models, we similarly iterate over loss functions and batch sizes but add quantum-specific parameters like the circuit ansatz (visualized in Fig.~\ref{fig:circuit_1} and Fig.~\ref{fig:circuit_2}), parallel encoding, and output scaling. This thorough search aims to capture how each hyperparameter choice influences predictive accuracy and optimization success, ultimately identifying robust configurations for each modeling paradigm. 

The hyperparameter cubes for the classical and quantum models are detailed in Table~\ref{tab:hyperparameters_combined}.

\begin{table}[tb]
    \caption{Hyperparameter Cubes for Classical and Quantum Models}
    \begin{center}
    \scriptsize
    \begin{tabular}{|l|l|l|}
    \hline
    \textbf{Hyperparameter}  & \textbf{Classical Values}                   & \textbf{Quantum Values}                                \\
    \hline
    \textbf{Number of layers}    & 1, 2, 3, 4                                   & 20, 40, 60                                             \\
    \textbf{Hidden sizes}        & 40, 64, 128                                 & \emph{N/A}                                            \\
    \textbf{Activation}          & \texttt{relu}, \texttt{tanh}, \texttt{sin}  & \emph{N/A}                                            \\
    \textbf{Batch size}          & 16, 32                                      & 16, 32                                                \\
    \textbf{Loss functions}      & MSE, MAE                                    & MSE, MAE                                              \\
    \textbf{Optimizers}          & Adam, SGD                                   & Adam                                                  \\
    \textbf{Input ranges}      & \([-1.0,\,1.0]\), \([0.0,\,1.0]\)           & \([0,\,1], [-1,\,1], [-0.5,\,0.5]\)               \\
    \textbf{Output range}        & \([-0.5,\,0.5]\)                            & \([-0.5,\,0.5]\)                                      \\
    \textbf{Stratify data}       & True, False                                 & True, False                                           \\
    \textbf{Sample weighting}    & True, False                                 & True, False                                           \\
    \textbf{Log scaling}         & True                                        & True                                                  \\
    \textbf{Circuits}            & \emph{N/A}                                  & 1, 2                                                  \\
    \textbf{Parallel encoding}   & \emph{N/A}                                  & True, False                                           \\
    \textbf{Output scaling}      & \emph{N/A}                                  & True, False                                           \\
    \hline
    \end{tabular}
    \label{tab:hyperparameters_combined}
\end{center}
\end{table}

\section{Results} \label{sec:results}

\subsection{Regression Accuracy} \label{subsec:regression_accuracy}

The top 10 best-performing models and their configurations for both quantum and classical models were identified based on validation MSE and ROG. To ensure robustness, each selected configuration was retrained 10 times using different random seeds but with identical hyperparameters, enabling an assessment of training uncertainties. 

\begin{figure*}[tb]
    \centerline{
    \begin{adjustbox}{width=1.0\textwidth}
        \begin{tabular}{cc}
            \includegraphics[width=0.5\textwidth]{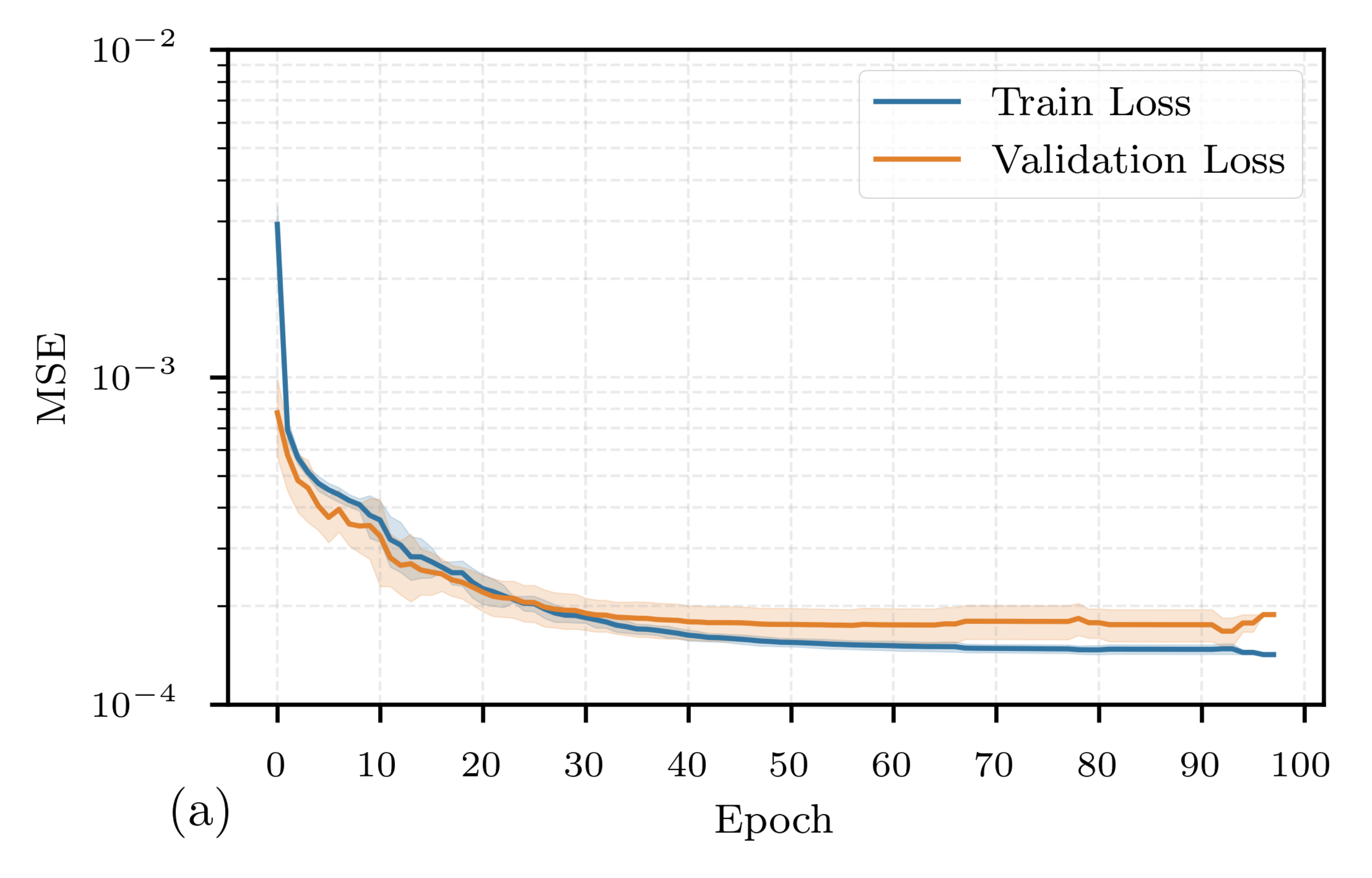} &
            \includegraphics[width=0.5\textwidth]{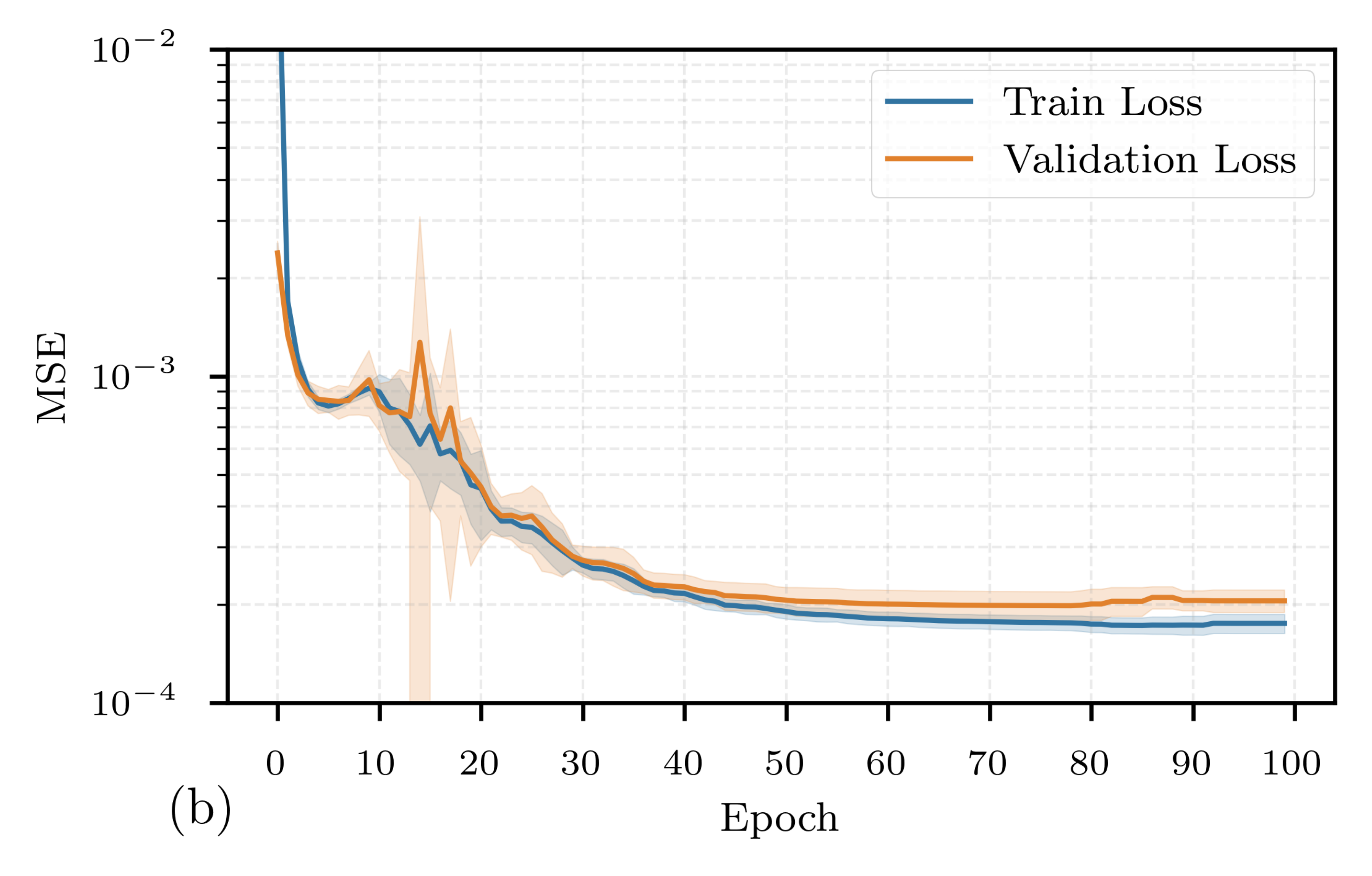}
        \end{tabular}
    \end{adjustbox}}
    \caption{Training and validation loss curves for the top 10 classical and quantum models. The shaded area around each line represents one standard deviation from the mean over ten repeated training runs with identical hyperparameters but different random seeds.}
    \label{fig:loss_curves}
\end{figure*}

Fig.~\ref{fig:loss_curves} shows the training and validation loss curves for a classical and quantum model configuration selected from the top 10 models based on minimum MSE and maximum ROG. From these, the configuration with the lowest average validation loss over 10 repeated training runs was selected for comparison.
The shaded area around each line represents one standard deviation from the mean (solid line) over ten repeated runs with identical hyperparameters but different random seeds. 
As evident in both figures, all four configurations converge to low validation losses, though the quantum models require more iterations and typically exhibit higher variance during initial epochs. 

Table~\ref{tab:best_configs} summarizes the best configurations for both classical and quantum models, including the number of parameters, average training time, validation MSE, and test MSE over 10 repeated training runs.
The classical models exhibit slightly lower average validation and test MSE values compared to their quantum counterparts. However, the quantum models demonstrate competitive performance, with the best quantum model achieving a test MSE of $2.066 \times 10^{-4}$, only marginally higher than the best classical model's $1.877 \times 10^{-4}$. The training times for quantum models are significantly higher than those for classical models due to the increased computational complexity of quantum circuits.

\begin{table}[tb]
    \centering
    \caption{Training Statistics Summary}
    \scriptsize
    \begin{tabular}{|l|c|c|c|c|}
    \hline
    \textbf{Model} & \textbf{\#Param.} & \textbf{Time} & \textbf{Val. MSE} & \textbf{Test MSE} \\
    \hline
 Classical (MSE) & 5359  & 1904 s & $1.723 \times 10^{-4}$ & $1.877 \times 10^{-4}$ \\
 Classical (ROG) & 4801  & 4801 s & $1.731 \times 10^{-4}$ & $1.889 \times 10^{-4}$ \\
 Quantum (MSE) & 968 & 30176 s & $1.914 \times 10^{-4}$ & $2.066 \times 10^{-4}$ \\
 Quantum (ROG) & 1448 & 65165 s & $1.980 \times 10^{-4}$ & $2.071 \times 10^{-4}$ \\
    \hline
    \end{tabular}
    \label{tab:best_configs}
\end{table}

\subsection{Optimization Performance} \label{subsec:optimization_performance}

In our experiments, we assessed the ability of both classical and quantum reward models to guide the PSO process toward superior steering configurations in the IB environment. Rather than simply relying on grid-sampled dataset values, the PSO process exploited the models' predictive power to explore the continuous action space for configurations yielding higher rewards.

The optimization pipeline uses trained surrogate models within PSO to refine candidate configurations per setpoint. These candidates are evaluated in the IB environment over multiple trajectories to obtain reliable ground truth fitness values, with improvements measured against the best available dataset configurations using the ROG metric.

\begin{figure*}[tb]
    \centerline{\includegraphics[width=1.0\textwidth]{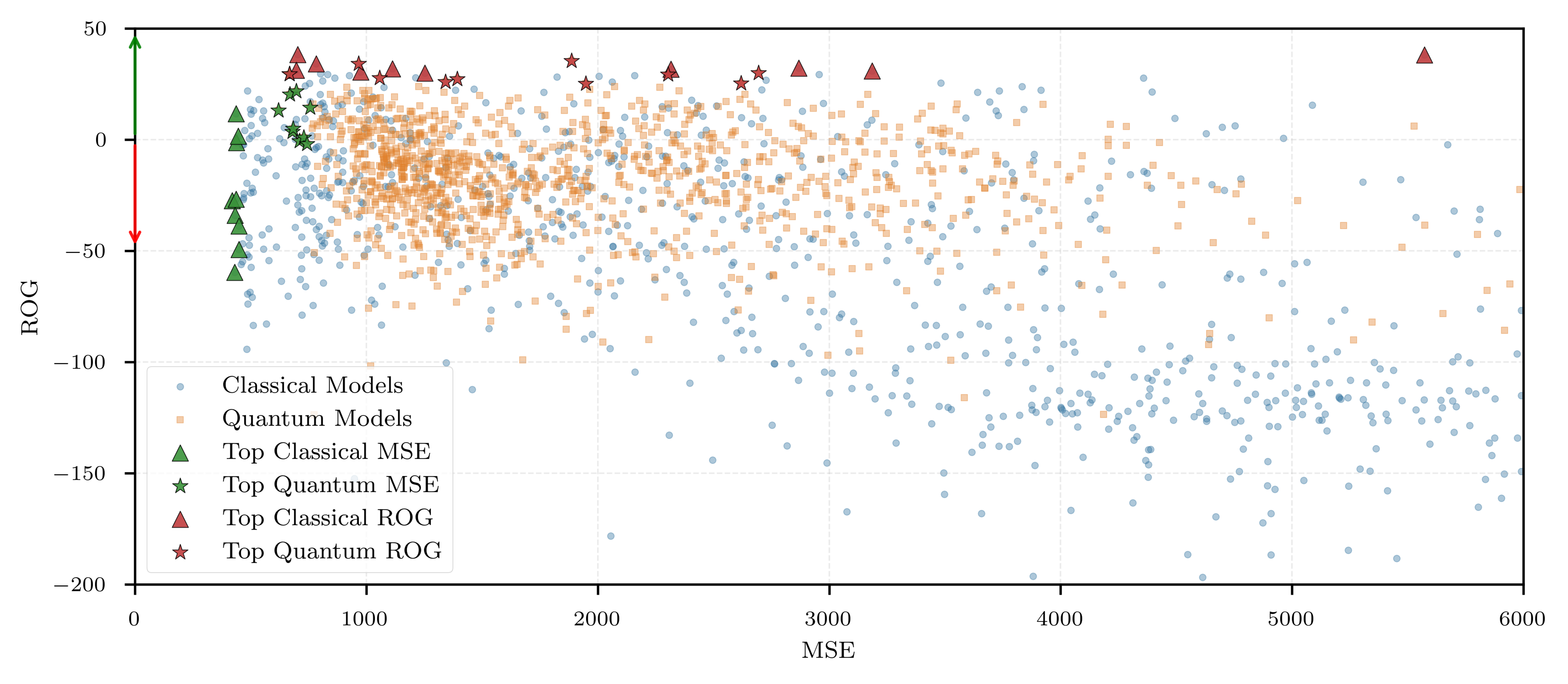}}
    \caption{MSE vs. ROG metric for all configurations in the hypercubes (zoomed-in region). For clarity, some models with large MSE values lie outside the displayed range. This closer view shows that certain classical models do reach slightly lower MSE values than quantum models. Nevertheless, a larger fraction of quantum configurations (263 out of 1152) surpasses the baseline dataset performance (ROG\textgreater0) compared to classical models (133 out of 1152). The green and red arrows mark regions of favorable and unfavorable performance.}
    \label{fig:MSE_vs_ROG}
\end{figure*}

Fig.~\ref{fig:MSE_vs_ROG} visualizes the relationship between regression accuracy (MSE) and optimization performance (ROG) for both classical and quantum model configurations explored in this study. Although some classical models exhibit lower MSE values compared to quantum models, it is noticeable that quantum configurations more frequently achieve positive optimization gains (ROG\textgreater0). Specifically, among all tested configurations, a larger number of quantum models (263 out of 1152) surpass the dataset baseline compared to classical models (133 out of 1152). This suggests that while lower prediction error is beneficial, it does not necessarily translate directly into superior optimization results. Quantum models, despite generally higher regression errors, appear more robust in achieving positive improvements over the baseline in this specific experimental setup.

Hence, this discrepancy motivates our choice to rely primarily on the ROG metric for model selection in the following figures.
ROG directly captures the practical value of models in an optimization context---quantifying tangible improvements over the dataset baseline---and thus provides a more meaningful criterion for identifying models that excel at real-world optimization tasks.

\begin{figure*}[tb]
    \centerline{
    \begin{adjustbox}{width=1.0\textwidth}
        \begin{tabular}{cc}
            \includegraphics[width=0.5\textwidth]{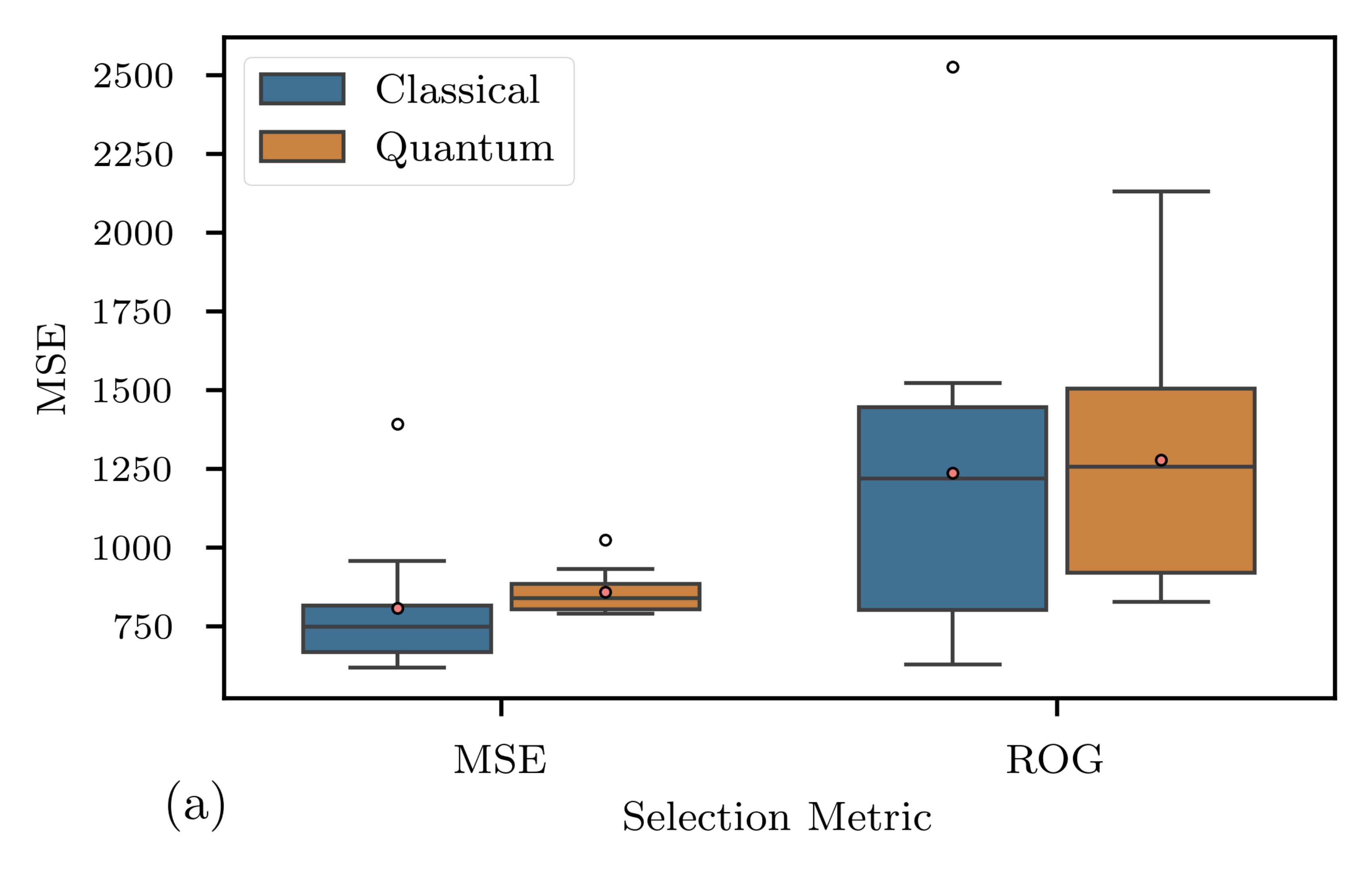} &
            \includegraphics[width=0.5\textwidth]{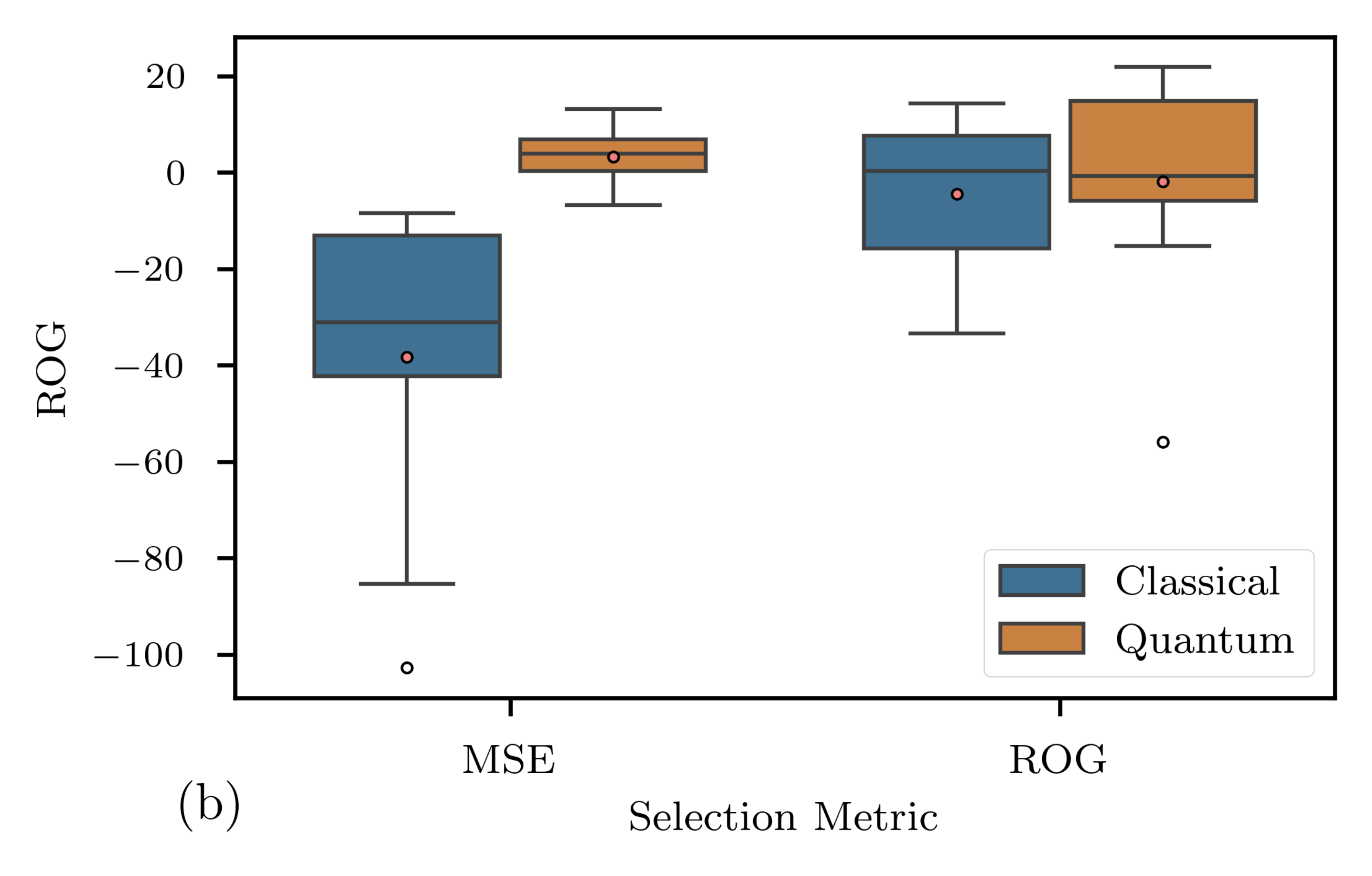}
        \end{tabular}
    \end{adjustbox}}
    \caption{(a) MSE value distribution for the top 10 classical and quantum models across all setpoints. (b) ROG value distribution for the top 10 classical and quantum models across all setpoints. The top 10 models were selected based on two criteria: minimum MSE and maximum ROG. The mean across the top 10 models is shown by the red marker. These aggregate metrics are further broken down by setpoint in Fig.~\ref{fig:rog_1x2_imp}.}
    \label{fig:top10_combined}
\end{figure*}

Fig.~\ref{fig:top10_combined} presents the distribution of the MSE and ROG for the top 10 classical and quantum models across all setpoints. 
In subplot (a), which displays the MSE distribution, classical models exhibit a wider range of MSE values, with some models achieving lower errors compared to their quantum counterparts. However, quantum models demonstrate more consistent performance, as indicated by their tighter interquartile range. Notably, models selected based on the ROG metric show greater variability in both classical and quantum configurations.
In subplot (b), the ROG distribution highlights the optimization effectiveness of both model types. Classical models selected based on MSE exhibit a broader spread of ROG values, including negative values, indicating that some models fail to discover improved configurations over the database baseline. Conversely, quantum models maintain higher and more stable ROG values, particularly when selected using the ROG criterion. This suggests that while classical models may achieve lower prediction errors, quantum models are better at generalizing beyond training data, leading to the discovery of superior configurations.

\begin{figure*}[tb]
    \centerline{\includegraphics[width=1.0\textwidth]{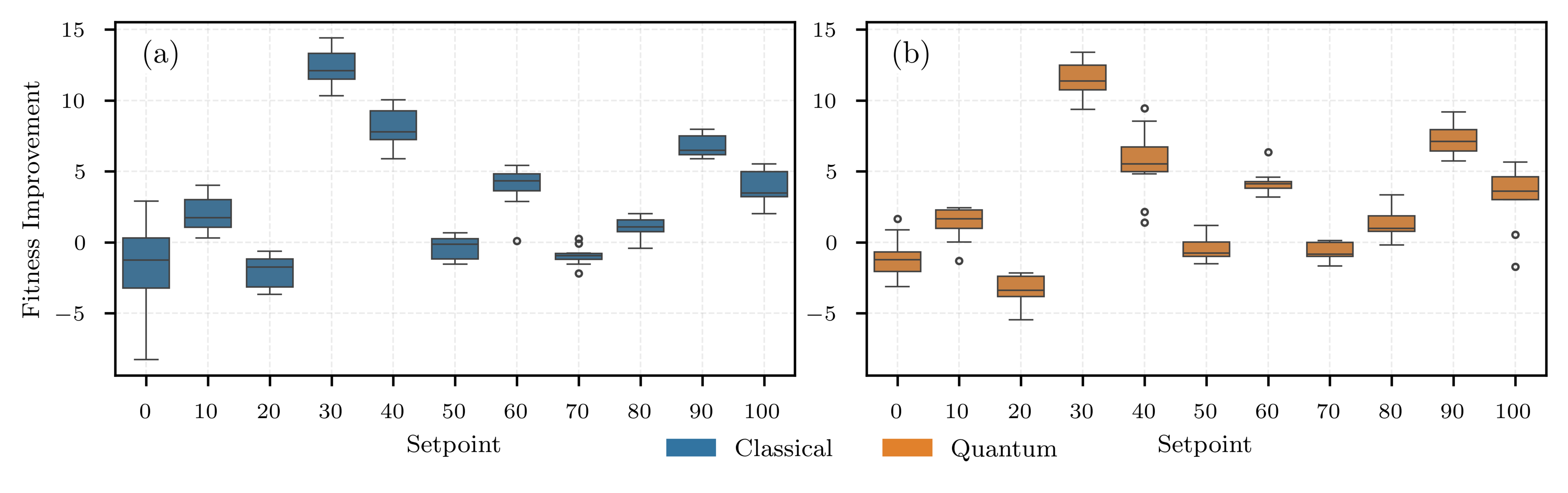}}
    \caption{Setpoint-wise breakdown of the optimization performance (ROG) shown in Fig.~\ref{fig:top10_combined}(b). Subplots (a) and (b) show the distribution of ground truth fitness improvements across individual setpoints for classical and quantum models, respectively. This detailed view reveals that both model types achieve peak improvements around setpoints 30 and 90, though some negative values where optimization fails to improve upon the dataset baseline are evident at intermediate setpoints.}
    \label{fig:rog_1x2_imp}
\end{figure*}

Fig.~\ref{fig:rog_1x2_imp} presents a setpoint-wise breakdown of fitness improvements for the \emph{top 10 models} selected by their ROG performance, complementing the aggregate ROG distribution shown in Fig.~\ref{fig:top10_combined}(b). Subplots (a) and (b) show how the top 10 models from each hypercube (classical and quantum) perform across the full range of setpoints, revealing significant variation in improvement trends. Notably, both model types achieve peak improvements around setpoints 30 and 90, though some negative values where optimization fails to improve upon the dataset baseline are evident at intermediate setpoints. 

These results suggest that while the best classical models from our hyperparameter search space achieve slightly higher average gains, the quantum models---despite significantly higher training times (see Table~\ref{tab:best_configs})---demonstrate competitive performance and greater consistency in the noisy, sparse data scenario. 
This is evident in both the aggregate metrics shown in Fig.~\ref{fig:top10_combined} and the setpoint-wise breakdown in Fig.~\ref{fig:rog_1x2_imp}, which together provide a comprehensive view of model performance across different scales of analysis. 
This suggests that quantum models can effectively guide the PSO search to identify superior configurations.

To further understand these performance differences, we analyze the impact of hyperparameter choices on both MSE and ROG metrics in Appendix~\ref{app:hyperparameter_influence}. This analysis reveals which architectural configurations consistently appear in the highest-performing classical and quantum models, providing additional insight into the model characteristics that lead to successful optimization outcomes.

\section{Discussion} \label{sec:discussion}

\subsection{Main Findings} \label{subsec:main_findings}

This work demonstrates the viability of VQCs as regression models for offline contextual bandit problems in an industrial setting, represented by the IB environment. Our results show that VQCs can effectively fit complex, non-linear reward functions derived from sparse, noisy industrial data. Critically, VQCs, when coupled with PSO, successfully identified steering configurations yielding higher rewards than those present in the original, grid-sampled dataset. This improvement is quantified by the positive ROG achieved by the best quantum models ($35.45$), which is competitive with the best classical models ($38.37$).

While classical models achieved slightly lower MSE in some cases (see Table \ref{tab:best_configs}), quantum models exhibited a greater tendency to achieve positive ROG (Fig.~\ref{fig:MSE_vs_ROG}), suggesting better generalization capabilities beyond the training data distribution. 
The success of VQCs hinges on appropriate circuit design choices. Parallel encoding, deeper circuit architectures, and trainable output scaling were found to be beneficial, corroborating findings in prior QML research \cite{schuldEffectDataEncoding2021, jerbiParametrizedQuantumPolicies2021, simExpressibilityEntanglingCapability2019a}. These design elements likely enhance the expressivity and trainability of the VQCs, allowing them to capture the complex relationships within the IB environment.

\subsection{Limitations and Challenges} \label{subsec:limitations}

Despite the promising results, several limitations and challenges must be acknowledged. A key constraint is the computational cost associated with training VQCs. As shown in Table \ref{tab:best_configs}, training times for quantum models were significantly longer than for classical NNs. While bigger circuits can be beneficial for convergence, simulating them is expensive and restricts experimentation, especially for current NISQ devices that suffer from noise and gate errors that make high-depth circuits impractical on real hardware. This limits the scale of VQC architectures that can be practically explored.

In addition, the current study relies on simulations of quantum circuits. Deployment on real quantum hardware, particularly in the current NISQ era, would introduce additional challenges related to noise, limited qubit connectivity, and gate fidelity.

\subsection{Future Work} \label{subsec:future_work}

Future research should focus on addressing the identified limitations and exploring the full potential of QML in industrial optimization. Two promising directions include:

\paragraph{Quantum Search Algorithms} 
Quantum search algorithms like Grover's algorithm offer theoretical speedup for optimization problems~\cite{baritompaGroversQuantumAlgorithm2005}. Having a precise model of the input-output relationship on a quantum computer could enable significantly accelerated identification of optimal configurations for each context through iterative probability amplification. However, applying this to our contextual bandit setting presents unique challenges distinct from recent work in quantum reinforcement learning. \citeauthor{wiedemann2023quantum}~\cite{wiedemann2023quantum} demonstrated quantum speedup for policy iteration in finite Markov decision process (MDPs) by combining amplitude estimation and Grover search, focusing on discrete state-action spaces where complete policies can be encoded into quantum circuits. In contrast, our contextual bandit setting would require performing quantum search directly over discretized action spaces for each individual context. This presents significant scaling challenges for industrial applications, as representing our three-dimensional continuous action space $(v,g,h)$ with sufficient precision would require an exponential number of qubits in the resolution of each action dimension.

\paragraph{Hardware Implementation} Recent developments in quantum hardware show promise in meeting these demanding requirements. Testing the proposed approach on real quantum hardware is crucial to assess its practical feasibility and identify the challenges associated with noise and hardware limitations. Recent advancements, such as Google's Willow chip with improved error correction capabilities~\cite{acharyaQuantumErrorCorrection2024} and Microsoft's exploration of topological qubits based on Majorana zero modes~\cite{aghaeeInterferometricSingleshotParity2025, aasenRoadmapFaultTolerant2025}, offer promising pathways towards more robust and scalable quantum computation. These advancements, while still in the early stages, could significantly mitigate the challenges posed by NISQ devices and make practical applications of VQCs more feasible.

\section{Conclusion} \label{sec:conclusion}

This work demonstrates the feasibility of using VQCs for offline contextual bandit problems, using the IB environment as a representative industrial optimization challenge.
We show that appropriately designed VQCs can effectively model complex reward functions and, integrated with PSO, discover superior configurations. Although practical limitations related to computational cost and current hardware exist, the strong generalization performance of VQCs, coupled with advances in quantum technology, makes them a promising avenue for tackling industrial optimization challenges and warrants further investigation in the field of QML.

\appendices

\section{Industrial Benchmark Environment} \label{app:ib_env}

\begin{figure}[tb]
    \centerline{\includegraphics[width=1.0\columnwidth]{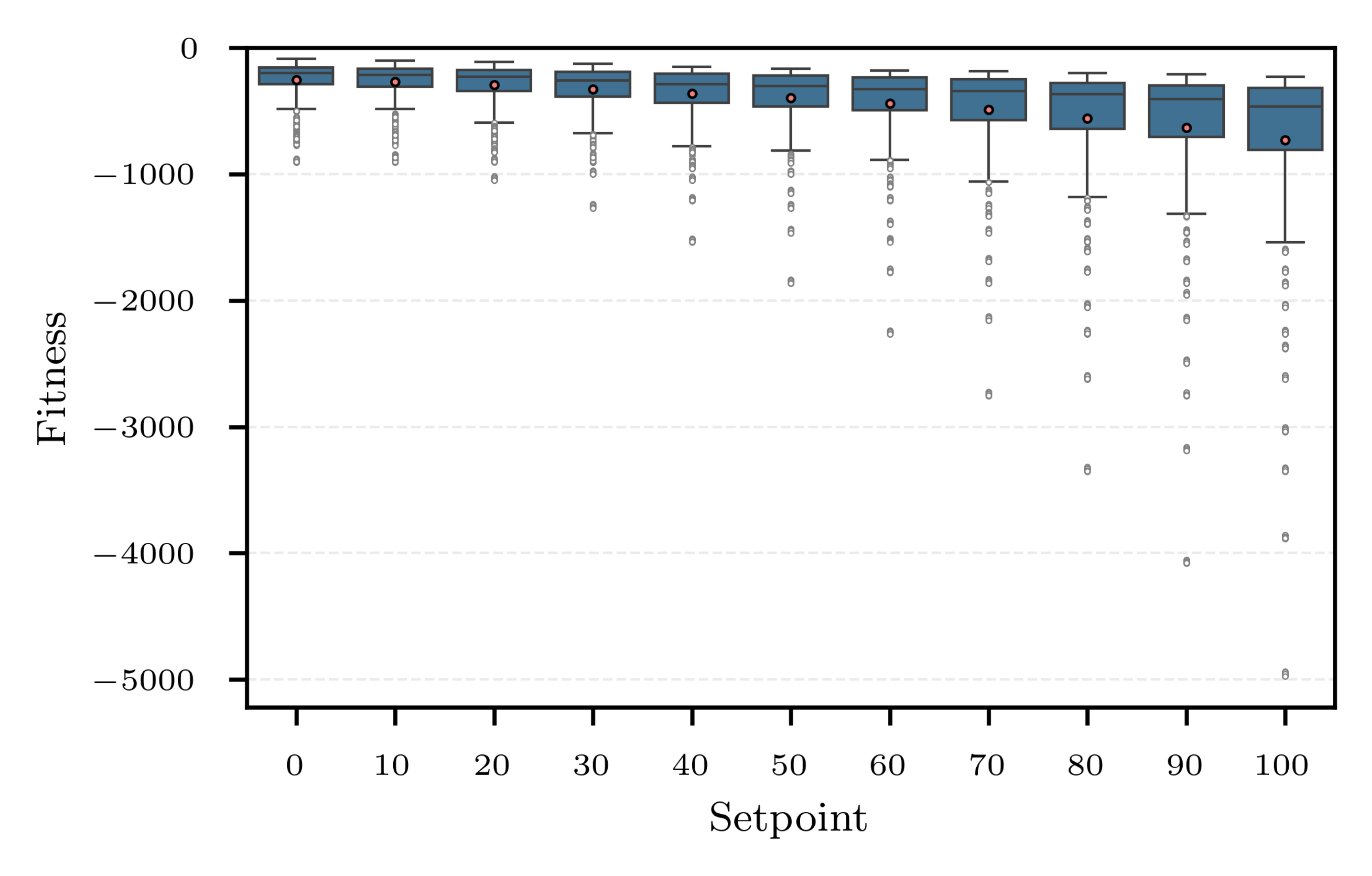}}
    \caption{Fitness value distribution per setpoint. Mean values are marked with a red dot. A visibly decreasing trend in fitness values can be observed as the setpoint increases.}
    \label{fig:ib_fitness_dist}
\end{figure}

The distribution of fitness values across setpoints in the IB environment is visualized in Fig.~\ref{fig:ib_fitness_dist}. The mean fitness values per setpoint are marked with red dots, showing a decreasing trend as the setpoint increases. Table~\ref{tab:fitness_statistics} provides a summary of the fitness value statistics per setpoint, including the mean, standard deviation, minimum, and maximum values. The fitness values exhibit a wide range, with the mean fitness decreasing from -253.49 at setpoint 0 to -729.27 at setpoint 100. The standard deviation increases with the setpoint, indicating greater variability in fitness values for higher setpoints.

\begin{table}[tb]
    \caption{Statistics of fitness values per setpoint.}
    \begin{center}
    \begin{tabular}{|c|c|c|c|c|}
    \hline
    \textbf{Setpoint} & \textbf{Mean} & \textbf{Std. Dev} & \textbf{Min} & \textbf{Max} \\
        \hline
 0   & -253.49 & 148.60 & -906.93 & -85.38 \\
 10  & -270.68 & 156.24 & -906.68 & -101.70 \\
 20  & -293.29 & 172.04 & -1048.88 & -113.11 \\
 30  & -327.83 & 200.06 & -1268.71 & -126.60 \\
 40  & -360.44 & 229.75 & -1537.14 & -148.79 \\
 50  & -395.78 & 271.06 & -1864.47 & -164.25 \\
 60  & -439.60 & 326.50 & -2265.11 & -178.99 \\
 70  & -491.41 & 398.74 & -2754.42 & -184.89 \\
 80  & -555.73 & 489.56 & -3353.20 & -198.87 \\
 90  & -631.31 & 602.98 & -4079.72 & -211.95 \\
 100 & -729.27 & 744.38 & -4972.63 & -229.21 \\
        \hline 
    \end{tabular}
    \label{tab:fitness_statistics}
    \end{center}
\end{table}

\section{Hyperparameter Influence on Model Performance} \label{app:hyperparameter_influence}

Understanding which hyperparameter configurations lead to the best-performing models is crucial for interpreting the observed trends in classical and quantum model performance. 

\begin{figure*}[htbp]
    \centerline{\includegraphics[width=1.0\textwidth]{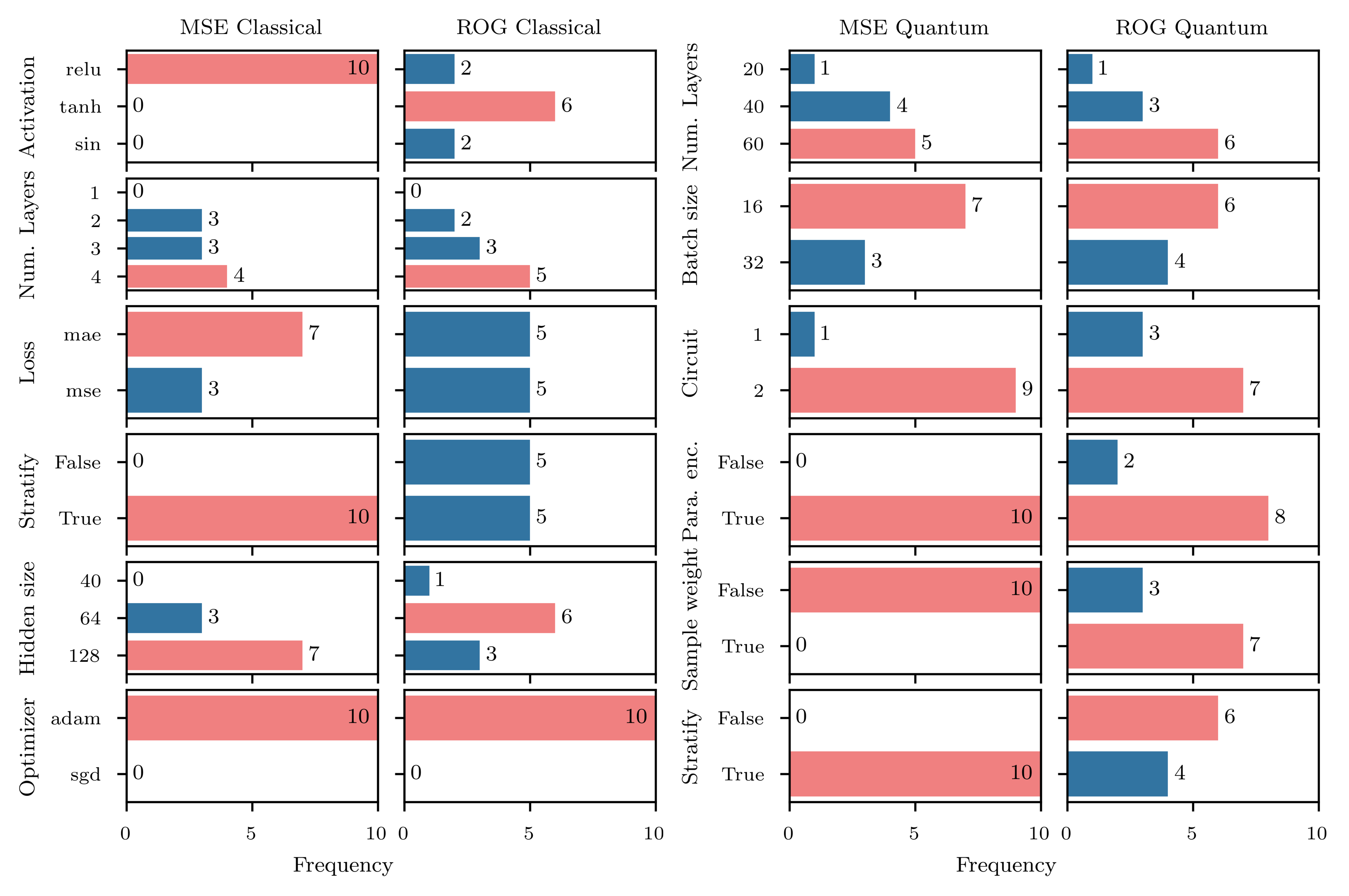}}
    \caption{Overview of the most frequent hyperparameter settings among the top 10 \emph{classical} models for each metric (MSE and ROG). Each column corresponds to a ranking criterion (MSE or ROG), and each row shows a particular hyperparameter. The bar heights indicate how many of the top 10 models share that hyperparameter choice, revealing trends such as a preference for ReLU and deeper architectures under MSE, as well as a broader mix of activation functions under ROG in classical models.}
    \label{fig:hyp_full_page}
\end{figure*}

Fig.~\ref{fig:hyp_full_page} provides an overview of a selection of hyperparameter settings that showed a clear preference among the top 10 classical models for each selection metric (MSE and ROG). 
Each row represents a specific hyperparameter, while the bar heights indicate the number of top-performing models that share that particular setting. This visualization reveals distinct preferences in hyperparameter choices based on whether models were optimized for minimizing MSE or maximizing ROG. For instance, classical models optimized for MSE tend to favor ReLU activation functions and deeper architectures, while those optimized for ROG exhibit a broader mix of activation functions. This analysis provides valuable insights into the hyperparameter configurations that lead to superior model performance in the contextual bandit optimization task.

\begin{figure*}[htbp]
    \centerline{\includegraphics[width=1.0\textwidth]{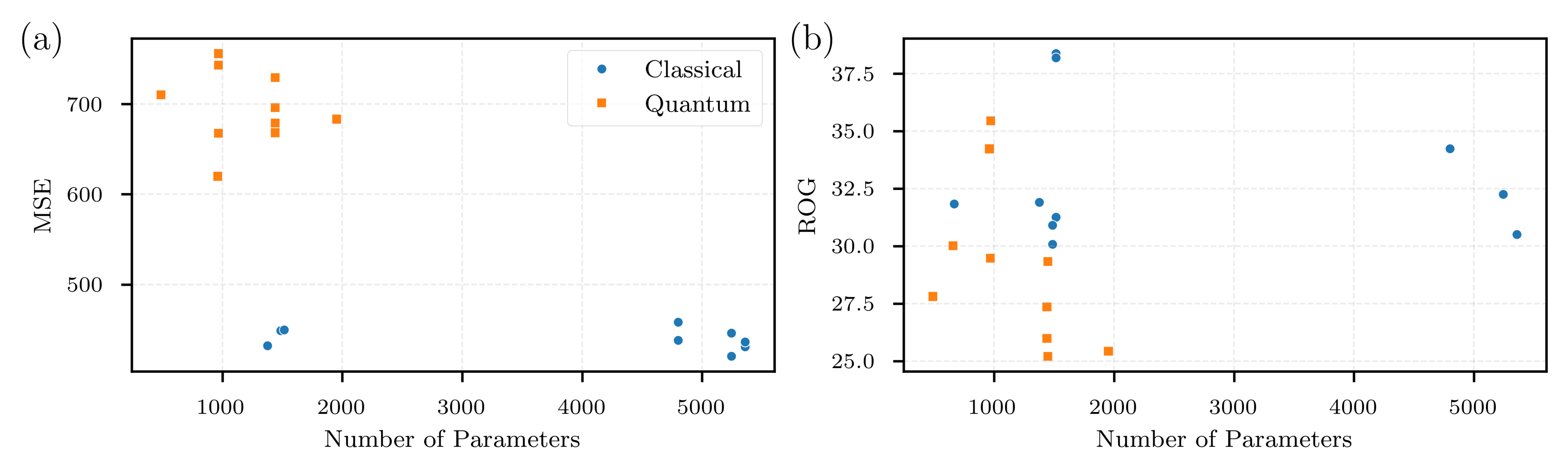}}
    \caption{Scatter plots comparing the top 10 classical and quantum models selected based on different performance metrics. The plots highlight the relationship between the number of parameters and (a) MSE and (b) ROG.}
    \label{fig:1x2_top10}
\end{figure*}

The relationship between model complexity (in terms of the number of parameters) and performance is further examined in Figure~\ref{fig:1x2_top10}, which shows how the top 10 classical and quantum models distribute across MSE and ROG metrics relative to their parameter count. Classical models tend to have significantly more parameters compared to quantum models, particularly in configurations optimized for MSE. However, quantum models achieve competitive ROG values despite their lower parameter counts, reinforcing the notion that quantum circuits can leverage expressivity more efficiently than classical networks.

Overall, the hyperparameter trends suggest that classical models benefit from depth, structured activation functions, and stratified data sampling, particularly when optimizing for regression accuracy. In contrast, quantum models favor deeper circuits, parallel encoding strategies, and more flexible data handling, particularly when optimizing for optimization performance (ROG). These findings highlight the distinct ways in which classical and quantum models navigate the trade-offs between accuracy, generalization, and optimization effectiveness in the contextual bandit setting.

\printbibliography

\end{document}